\documentclass[prl,twocolumn,floatfix,showpacs,superscriptaddress]{revtex4-1} %

\usepackage{amssymb}
\usepackage{graphicx}
\usepackage[shortlabels]{enumitem}
\usepackage{mathptmx} 

\begin{document}

\title{The Superconducting Plateau }

\author{Yaron Kedem}
\affiliation{Department of Physics, Stockholm University, AlbaNova University Center, 106 91 Stockholm, Sweden}
\email{kedem@kth.se}

\begin{abstract}
Two recent experiments in lightly doped strontium titanate have shown that the superconducting critical temperature remains constant in a range of carrier concentrations covering almost three orders of magnitude. The importance of this phenomenon, hereby dubbed a superconducting plateau, is comparable to the celebrated superconducting dome. The dramatic empirical evidence impose severe constraints on any proposal for a superconducting mechanism at low doping. We analyze the experimental results to see what type of theoretical models can be suitable in this regime. We show that the low critical current, absence of diamagnetic screening and the high critical field are all related to the small momentum of the carriers, which is expected at low densities. We quantify the dependency of the critical temperature on density by introducing the ``density exponent''. We discuss the role of a quantum critical point in the formation of a plateau and comment on the need for more experimental data.

\end{abstract}

\maketitle
Superconductivity in low doping systems, and in particular strontium titanate(STO), is a standing mystery for  more than half a century\cite{sto1a,sto1}. While there is a wide agreement that the conventional theories, namely BCS\cite{bcs} and its extensions to strong coupling\cite{migdal,eliashberg}, are not applicable, since many of the necessary assumptions are not valid in this regime and many of their predictions are false, there is no consensus on any alternative theory. Two recent experiments\cite{diluteAnand, rischau} placed a crucial constraint to the nature of such theory. They showed that the critical temperature, $T_c$, is essentially constant for a range of carrier densities, $n$, covering almost three orders of magnitude. This vast plateau was predicted in \cite{novel} but it is in a stark contradiction with almost any other theory of superconductivity.

A widely celebrated feature arising from the dependency $T_c$ on the level of doping is the superconducting dome. The superconducting plateau shares some of the properties that made the dome such a prolific phenomenon for physics research. First, it is a distinct qualitative feature. While the numeric value of $T_c$ or the particular level of doping where it starts/ends might vary, the marked shape can be easily recognized. Second, it sets a heavy burden on theoretical models. The conflict with the excepted behavior, based on conventional theories, together with irrefutable experimental evidence that cannot be tuned or misinterpreted, imply numerical calculations would not suffice here and a new paradigm must rise. It might also be interesting to note that STO was (one of) the first material where a superconducting dome was observed. One can wonder whether more plateaus might appear if experimentalists would actively search for them.

In recent years, the long lasting discussion regarding the superconducting mechanism in STO, has shifted focus. Following early work on the quantum nature of its paraelectric phase \cite{quantPara,quantFluct} some experiments observed a zero temperature transition to a ferroelectric (FE) phase \cite{Itoh1999,rowley}, implying the existence of a quantum critical point (QCP). This motivated some authors to propose it is the accompanying FE soft modes that mediate superconductivity and predicted an unusual isotope effect \cite{JonaPrl,prbIso}. The experimental confirmation of this effect \cite{stucky,tomioka, rischau} is all but a smoking gun connecting superconductivity and FE. The controversy is not over due to the widespread notion that those soft modes are nothing but the transverse optical phonons, which hardly couple to charge carriers. Different authors have proposed different solutions to this conundrum \cite{TO,mech, SCSTOrev}. However, the problem does not exist if one note that close to a QCP and when the continuous spherical symmetry is explicitly broken to a discrete cubic or $D_4$, which is likely the case for STO, the longitudinal modes soften as well \cite{paradigm}.

A major challenge in the context of low doping superconductivity is the invalidity of the so-called adiabatic approximation,
\begin{equation}\label{adiab}
E_F \ngtr \omega_c,
\end{equation} where $E_F$ is the Fermi Eenergy and $\omega_c$ is the typical frequency of the phonons. The challenge is multifaceted with issues concerning technical, physical and phenomenological aspects of the problem. The technical side manifests in the introduction of $\omega_c$ as a natural cutoff that considerably simplify the derivation. Since it is rather hard to make any analytical progress without this simplification, many authors introduce a cutoff even when there is no clear justification. The physical picture of superconductivity relies often on retardation, where an electron is attracted to the positively charged wake left by another fast electron due to the slow moving ions. In the case of Eq. (\ref{adiab}), the ions can be quicker than the electrons and there is no obvious way to overcome the Coulomb repulsion. From a phenomenological perspective, the simple question is how would $T_c$, the critical current, the tunneling gap and other observed quantities depend on $E_F$, or rather on the chemical potential $\mu$, as $\mu \rightarrow 0$. 

In \cite{novel} it was shown that one can employ the properties of a quantum phase transition to solve these issues. The divergent correlation length associated with phase transitions implies extremely long range interactions that are then limited to tiny momenta. This scenario, sometimes called forward scattering, justifies introducing a delta function in momentum space, relieving us from the need for a cutoff while also simplifying the derivation. The vanishing energy scale at a QCP means vanishing frequencies for the associated modes so the system reacts infinitely slowly allowing retardation even for slow electrons. The model led to bold predictions: at vanishing levels of doping, close to a FE QCP, the critical temperature is independent of the chemical potential, i.e. the superconducting plateau that was observed in \cite{diluteAnand, rischau}.

\begin{figure}
\includegraphics[width=0.48\textwidth]{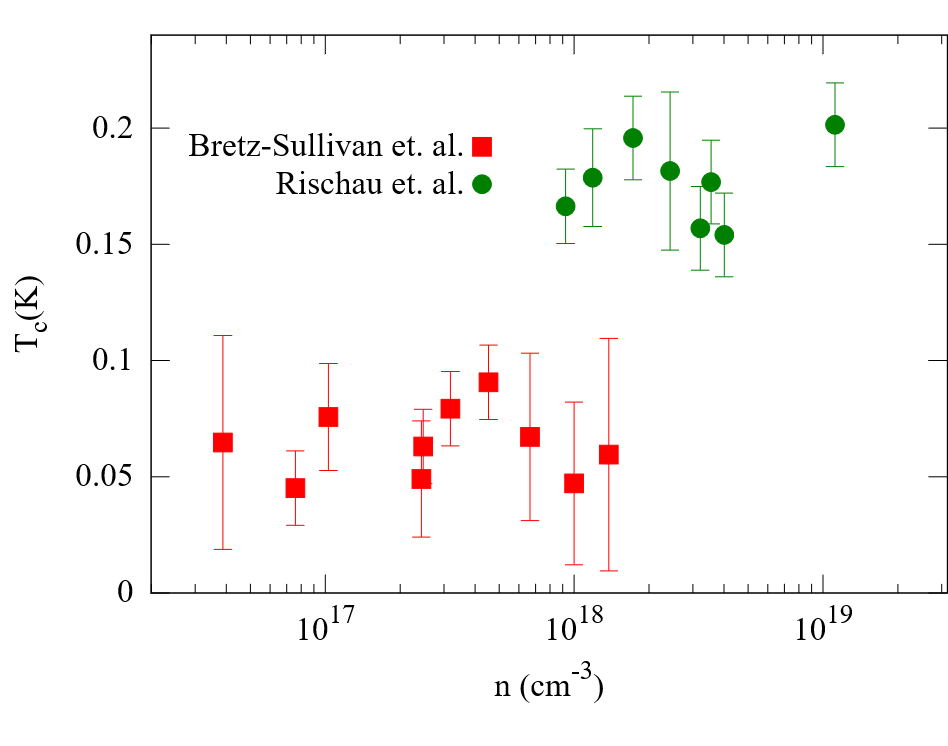}
\caption{The superconducting plateau. Experimental results from \cite{diluteAnand} and \cite{ rischau} for oxygen reduced samples and no other major modifications, such as isotope or calcium replacement, stress etc. The plateau is most evident in the data of \cite{diluteAnand} (red squares) and indeed the authors clearly stated “$T_c$ does not vary appreciably when $n$ changes by more than an order of magnitude”. The focus of \cite{ rischau} was on the superconducting dome, occurring at higher densities (not shown here) so the authors did not comment on the behavior at this regime. Nonetheless, their results (green circles) also show a range larger than an order of magnitude where $T_c$ varies by no more than 30\%, a variation that is comparable to the uncertainty in $T_c$, manifesting the same qualitative plateau. Any dome-like feature or some general trend one might want to infer from this data, would be extremely fine tuned and fragile. Considering the data from both groups together, the lower $T_c$ for lower densities seems to stem mainly from technical/methodological issues as the the difference persist even for samples with overlapping densities.   }
\label{exp}
\end{figure}

Fig. \ref{exp} shows the experimental evidence for the superconducting plateau. Measurements of $T_c$ in a range of densities covering almost three orders of magnitudes reveal no qualitative change. We focus on samples in which the density of carriers is the only proclaimed tuning parameter. Other published results, which are not shown here, do not alter the conclusion. These experimental results pose a major theoretical challenge. The challenge is not calculating the numerical value of $T_c$ but conceiving a theory where $T_c$ does not depend on the density of carriers. Any proposal for the superconducting mechanism for low doping systems and in particular STO must, at the very least, reproduce the plateau.

At the higher density side, the plateau ends at the start of the dome, roughly around $n \simeq 2 * 10^{19} cm^{-3}$. It is not clear how long it can stretch into lower densities. One possible boundary is  the intersection of  $E_F$ and $T_c$ as beyond this point the carriers do not form a degenerate Fermi Gas. In \cite{novel} it was suggested that thermally excited electrons can still bind via certain interactions but it is not clear if such bound states can form a condensate and allow a superconducting current. A naive approximation, based on free electrons at  $n = 10^{15} cm^{-3}$ yield $E_F \simeq 40 mK$, lower than the measured $T_c$ at the plateau. Even when $E_F > T_c$ it is possible that $E_F \simeq \Delta$, the superconducting gap as usually $\Delta > T_c$. The ratio $\Delta / T_c$ is 3.5 in BCS theory, in \cite{novel} it was derived to be 4, \cite{diluteAnand} estimate it to 2 -17 and \cite{measureGap} used microwave stripline resonators to measure values in agreement with BCS at higher densities. When $E_F < \Delta$ the superconducting gap and the ordinary band gap of the parent insulator are intermixed. More theoretical work is required to clarify what phenomenology is expected in this regime.

Let us now comment the relevance of percolation, weak-link, or disorder driven transition to the superconducting dome. In \cite{diluteAnand} the authors attempted to use percolative models to explain the width of the transition and obtained partial agreement with the data. Indeed it is likely that the transition has a percolative comportment, but it is important to note that percolation theory itself does not explain, let alone predict, the vast plateau. See Appendix for a more detailed discussion regrading these type of models.    

We look now more closely at the results of these experiments and their implications. The first takeaway from the plateau is that one has to abandon the Fermi surface paradigm, or more informally stated, it is not longer true that everything happens on the Fermi surface. Indeed, in elemental metal like iron with $E_F \simeq 10^5 K$, it is likely only the electrons closest to the Fermi surface participate in most relevant processes. However, already when we lose the adiabatic approximation (\ref{adiab}) we lose justification to neglect electrons from the bottom of the band. Nonetheless, quantities such as the Fermi velocity, Fermi momenta, $\vec{k}_F$ or the density of states at the Fermi surface, play a prominent role in many derivations that refer to non adiabatic scenarios. The vast superconducting plateau aggravate the issue as those quantities vary over orders of magnitude while the critical temperature remain constant.

We can demonstrate this point by looking on the estimation of the critical current density $j_c$. In \cite{diluteAnand} they have found a discrepancy of four orders of magnitudes between the measured  $j_c$ and the estimated   
\begin{equation}\label{depair}
j_{depairing} = e n {\Delta \over \hbar k_F},
\end{equation}
where $e$ is the elementary charge. The derivation of Eq. (\ref{depair}) is based on assuming a Galilean boost to a velocity $\vec{v}$ increases the energy of carriers with momentum $\vec{p} =\hbar \vec{k}$ by an additional contribution of $\vec{p} \cdot  \vec{v}$, which has to be smaller than $\Delta$. The derivation neglects another contribution to the energy $ {1 \over 2} m v^2 $, where $m$ is the (effective) mass of the carriers. This approximation is justified in many cases since the relevant carriers are those close to the Fermi surface  $ |\vec{k}| \simeq k_F$ and the boost velocity is small compared to the Fermi velocity $v_F = \hbar k_F / m  \gg v $. But not in our case. In fact when $\Delta \simeq E_F$ it is possible for carriers at the bottom of the conduction band, with vanishing momenta $k \rightarrow 0$, to still contribute to the critical current.

Taking into account $ {1 \over 2} m v^2 < \Delta  $  (and neglecting $\vec{p} \cdot  \vec{v}$) we obtain another estimate
\begin{equation}\label{depair2}
j_{depairing} = e n \sqrt{\Delta \over 2 m},
\end{equation}
which yield the correct order of magnitude. Note that we do not attempt to get a precise quantitative agreement with the measured results. For example, setting in Eq. (\ref{depair2})  $\Delta = 2 k_B T_c$ and $m=2 m_e$, with $m_e$ the electron mass, yields a $j_{depairing}$ that is larger by a factor of 2-4 than the measured results. It is hard to know, without further measurements, whether this factor is due to the neglected $\vec{p} \cdot  \vec{v}$ or simply because we did not use the correct values for $\Delta$ and $m$. 

Two other features discussed in \cite{diluteAnand} are high upper critical fields and an absence of diamagnetic screening. At first one might see these as contradicting since high critical field might be associated with a large gap and weak screening with a vanishing gap. However, both are explained by low velocity carriers. To gain intuition let us consider first classical electrodynamics. A slow moving charge produces weak magnetic filed, on the one hand, and is weakly affected by the Lorentz force, on the other hand. In quantum mechanics the coupling term between charge and field is $\vec{p} \cdot  \vec{A}$, where $\vec{p} $ is the momentum of the charge and $ \vec{A}$ is the vector potential. Small $\vec{p} $ of mobile electrons means they cannot efficiently screen magnetic fields and also that strong fields are required to modify their energy significantly. 

The superconducting dome is often associated with a QCP and in \cite{novel} it was shown that a QCP is indeed responsible to the appearance of a superconducting plateau. In contrast to the typical situation in High-$T_c$ superconductivity, in the case of STO one can study the QCP directly. Without introducing carriers STO is an insulator and by fabricating isotope-substituted $SrTi(^{18}O_y^{16}
O_{1-y} )^3$ samples one can tune the zero temperature phase to be paraelectric, for $y < 0.33$ or ferroelectric for $y > 0.33$. Naively one might think the atoms do not move at zero temperature, i.e. there are no phonos, so how would the atomic mass come in the equations and alter the ground state of the system? Indeed it seems that in the paraelectric ground state the ions are not in a well-defined symmetric position but rather in a superposition of two symmetry breaking positions, the so called ``quantum fluctuations''. At $y \simeq 0.33$ the paraelectric and ferroelectric states become degenerate so one can easily shift the system from one state to another. In this QCP the system is highly susceptible with extreme long range correlations, a most beneficial disposition for mediating superconductivity.

Introducing charge carriers complicates the picture. In a conductor one cannot directly measure the dielectric constant, whose divergence is the most distinct feature of a FE transition. In fact certain definitions of FE do not apply to conducting materials and often in the literature people refer to it as ``ferroelectric-like'' (FEL). The possible coexistence of superconductivity and FEL was shown in \cite{FEinDome} by substituting of a tiny fraction of strontium atoms with calcium. This method creates local fields at the substitution sites which can then percolate but do not necessarily bring the system closer to the QCP. In the case of isotope substitution, we know the location of the QCP without carriers but for a finite $n$ we do not know what value of $y$, if any, is required to make the ground state FEL.

\begin{figure}
\includegraphics[width=0.48\textwidth]{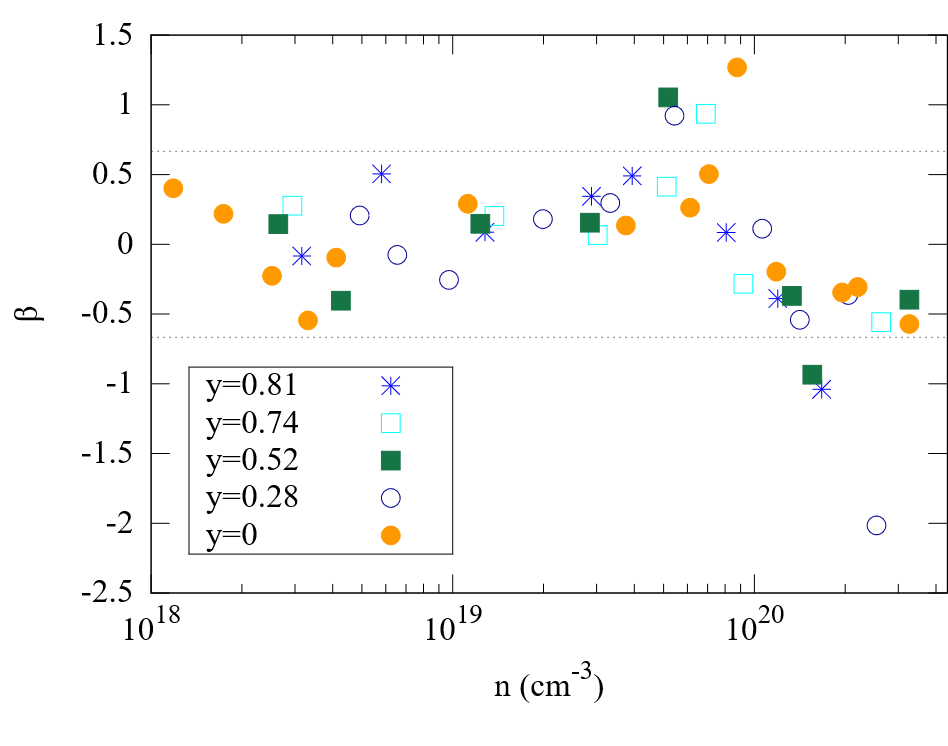}
\caption{The carrier density exponent. To estimate Eq. (\ref{denExp}) we plot a discrete derivative $\beta \simeq \Delta T / (T * \Delta \ln n)$ of the data presented in \cite{rischau}. Dashed lines at $\beta = \pm 2/3$ represent the case $T_c \propto E_F \propto n^{2/3}$. We see moderate values of $\beta \simeq \pm 1$ at the edge of the dome, $n \simeq  5*10^{19}, 2*10^{20} cm^{-3}$. For $n < 2 *10^{19} cm^{-3}$  $\beta$ is small and essentially fluctuate around 0.}  
\label{beta}
\end{figure}
In \cite{rischau} the parameter space $(n, y)$ was investigated extensively. Their focus was on the dome and indeed they confirmed the predicted shift to lower densities when $y$ is increased \cite{JonaPrl}. The shift, together with a broadening of the dome, reduces the extent of the plateau and somewhat blurs the boundary between the dome and the plateau. For $y= 0, 0.28, 0.52$ $T_c$ drops sharply at $n\simeq 8*10^{19}, 4*10^{19} ,3*10^{19} cm^{-3}$ respectively. For higher values of $y$, $T_c$ decreases more gradually with $n$. To quantify the dependency $T_c(n)$ we introduce, analogously to the isotope exponent, the carrier density exponent,    
\begin{equation}\label{denExp}
\beta = {\partial \ln T_c \over \partial \ln n}.
\end{equation}
Fig. \ref{beta} shows a discrete estimation of $\beta$ based on data from \cite{rischau}. In conventional theories, the minimal value $\beta=2/3$ can be obtain in case $T \propto E_F \propto n^{2/3}$, assuming three dimensional system with quadratic dispersion. We can see a wide range of densities with $|\beta| < 2/3$ that we consider as a plateau.

According to the model in \cite{novel} within the plateau the distance from the QCP is the main factor determining $T_c$. The location of the QCP, or rather the critical line this point traces out in $(n, y)$ parameter space, is not known. In \cite{rischau} the location of an anomaly in resistivity in this parameter space was measured. However this anomaly occurs even for $y<33$ where we do not expect a FEL transition. Raman spectroscopy of a transverse optical phonon for $y=0.47, n=0$ revealed it softens in a different temperature than the divergence of the dielectric constant, the most distinctive feature of a FE transitions. Thus while these measurements give us much information and provide circumstantial evidence for a FEL transition, is not likely we can use them to pinpoint the location of the QCP. 

\begin{figure}
\includegraphics[trim={0 0 1.5cm 2.5cm},clip,width=0.48\textwidth]{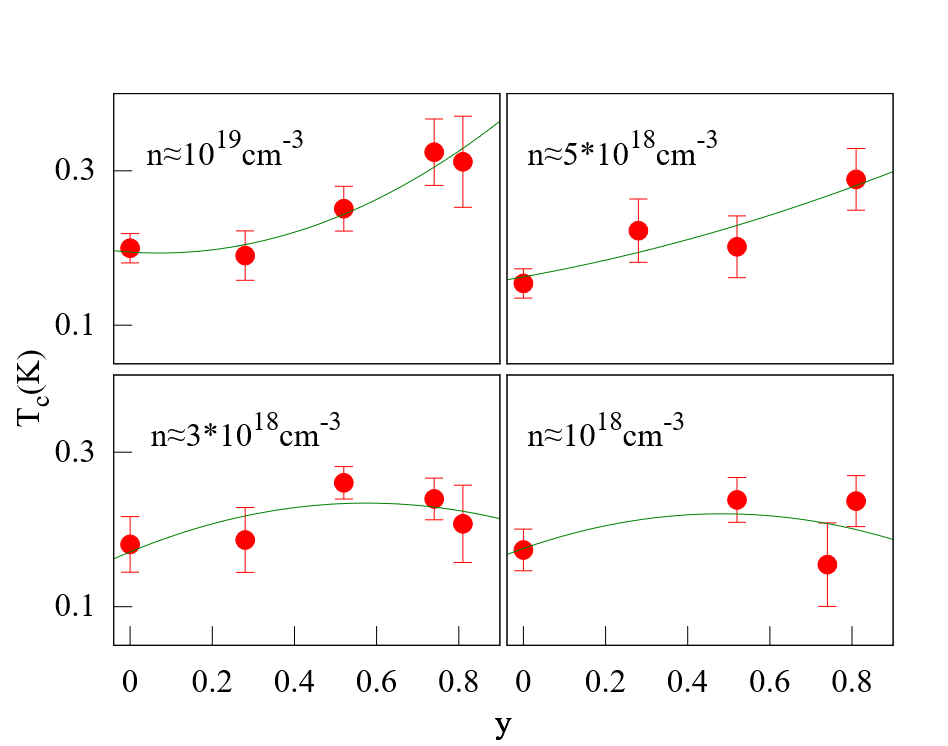}
\caption{Isotope effect in different carrier densities. All the data is taken from \cite{rischau}. Each panel shows results from samples with similar densities. Solid lines are a second degree polynomial fit. For high densities (top two panels and also for higher densities not shown here) one can see a monotonous enhancement of superconductivity with increasing $y$. For lower densities (bottom two panels) the effect seems to saturate or even reverse. It is possible that samples with small $n$ and and high $y$  are already in a FEL phase at the relevant temperatures. In these phase increasing $y$ would move the system further away from the QCP and suppress superconductivity.}
\label{iso}
\end{figure}
We attempt to gain further insight into this issue by looking on the dependency of $T_c$ on $y$ for different $n$, as shown in Fig. \ref{iso}. The negative trend of $T_c(y)$ for $n<5*10^{18} cm^{-3}, y>0.5$ is another circumstantial evidence that the system is in a FEL phase in this parameter subspace. Unlike the existence of the plateau, which is based on tens of samples in a wide range, this conclusion is somewhat speculative and relies only on a handful measurements. Thus, while currently the plateau is mostly a theoretical challenge, the problem of locating the QCP seems to lie at the experimentalists doorstep.
   
We have analyzed the recent experimental observation of a superconducting plateau in STO. This remarkable phenomenon opens a new research avenue where theoretical models can be tested experimentally. The experimental observation is a strong support for the model and the assumptions that led to the prediction of the plateau in \cite{novel}. Any other model or theory for superconductivity in this regime would have to reproduce the plateau to be considered. We have shown that many of the widespread methods are not applicable in this regime and physicists have to adopt new ways of thinking to approach the relevant problems. Many questions remain open: What are the boundaries of the plateau? Can we enhance superconductivity by moving closer to a QCP? Will we find similar plateaus in other materials? 

\begin{acknowledgments}
I would like to thank D. van der Marel for useful discussions
\end{acknowledgments}

\begin{widetext}

\appendix
 



\section*{Appendix: Ruling Out Percolation}

Let us first clarify the issue we are going to discuss in this section. The question is not whether there is a percolation process in the transition to superconductivity, but whether such a model can explain the plateau itself, i.e. can we use disorder to obtain Tc that is independent of doping? Or alternatively should we see the plateau as evidence for a percolation transition? While the answer seems to be negative, the arguments might suffer from a straw man fallacy, since no one has formally presented any such explanation. However, the notions of percolation, disordered or weak-link transitions are often mentioned in less formal settings such as private communications, seminars or as side remarks in the literature. Thus, it is constructive to constraint this narrative by analyzing its validity.

The general idea is that even above $T_c$ there are already superconducting regions and the transition occur when they are sufficiently connected to transfer charge across the sample. From the start we see that the superconducting plateau is in contrast to the usual percolation scenario, which is typically tuned by density, or characteristic length, and not by temperature. Nonetheless, one can still imagine that somehow the connectivity of the regions is controlled by the temperature and not so much by the density of carriers. For this to be the case, we need three conditions to occur:

\begin{enumerate} [(a)]
  \item The onset of superconductivity inside the regions at temperature higher than the measured Tc all across the plateau. \label{c1}
  \item  The properties of the regions, such as size or density, to be largely independent of doping. \label{c2}
  \item  A strong temperature driven effect that connects the regions at $T_c$. \label{c3}
  \end{enumerate}

To obtain \ref{c1} we should consider $T_c^{intra}(n)$, the onset temperature of superconductivity inside the regions. $T_c^{intra}(n)$ is unobserved since the regions are unconnected above $T_c$. We still require  $T_c^{intra}(n= 4* 10^{16} cm^{-3}) >60mK$ since sample with this density become superconducting at this temperate. Now the question is $T_c^{intra}(n= 10^{19} cm^{-3}) = ?$. The usual exponential dependency might give us room temperature. Even the “strong coupling” sub linear dependency $T_c \propto E_F \propto n^{2/3}$ would yield a sizable temperature.

Alternatively one might argue that carrier density inside the regions is constant due to some clustering in the doping procedure. This is at odds with the experimental observation of a homogeneous Fermi gas. In any case the density of such clusters would vary strongly and it seems unlikely this will have no effect on the measured Tc. 

It is hard to infer the implication of \ref{c2} without a particular model for the appearance of these regions. One would assume they could be described by quantities such as coherence length, penetration depth etc., which typically depend on the density. For example, if one would consider clusters of doping sites, it is hard to imagine that neither the density inside the cluster, nor the density of clusters would be affected by the overall density.

 To obtain \ref{c3} one might postulate a variety of phases or microscopic effects. It is not possible to completely rule out some drastic change in the material, unrelated to the presence of carriers, occurring at $T_c$ and connecting the isolated regions. However it seems unlikely that such transformation can only be observed via the onset of superconductivity. Unless the additional effect can be seen independently, we should apply Occam razor and discard it.
 
\end{widetext}

\end{document}